\documentclass[twocolumn,amsmath,amssymb,aps,prl,10pt]{revtex4-1}
\usepackage{graphicx}
\usepackage{dcolumn}
\usepackage{bm}
\usepackage{textcomp}
\usepackage[dvipsnames]{xcolor}

\newcommand{\ms}{ms${}^{-1}$\,}

\begin{document}
\title{Intermediate-range Casimir-Polder interaction probed by
high-order slow atom diffraction}

\author{C.~Garcion$^1$, N.~Fabre$^1$, H.~Bricha$^1$, F.~Perales$^1$,
S.~Scheel$^2$, M.~Ducloy$^{1}$, and G.~Dutier$^1$}
\email{dutier@univ-paris13.fr}
\affiliation{$^1$ Universit\'{e} Sorbonne Paris Nord, Laboratoire
de Physique des Lasers, CNRS, (UMR 7538), F-93430, Villetaneuse, France}
\affiliation{$^2$ Institut f\"ur Physik, Universit\"at Rostock,
Albert-Einstein-Stra{\ss}e 23-24, D-18059 Rostock, Germany}

\begin{abstract}
At nanometer separation, the dominant interaction between an atom and a
material surface is the fluctuation-induced Casimir--Polder potential.
We demonstrate that slow atoms crossing a silicon nitride transmission
nanograting are a remarkably sensitive probe for that potential. A 15\%
difference between nonretarded (van der Waals) and retarded Casimir--Polder
potentials is discernible at distances smaller than 51 nm. We discuss the
relative influence of various theoretical and experimental parameters on the
potential in detail. Our work paves the way to high-precision measurement of
the Casimir--Polder potential as a prerequisite for understanding fundamental
physics and its relevance to applications in quantum-enhanced sensing.
\end{abstract}

\maketitle

The Casimir--Polder (CP) interaction between an atom or molecule and
polarizable matter \cite{CasimirPolder1948} has been intensively studied
theoretically as a fundamental electromagnetic dispersion force
\cite{Scheel2008,Buhmann2012}. It originates from quantum fluctuations of the
electromagnetic field that spontaneously polarizes otherwise neutral objects.
Interaction strength and spatial dependence are the result of a unique
combination of atom species, internal atomic state and material properties and
geometry. The Casimir--Polder interaction is part of a larger family of
fluctuation-induced electromagnetic forces that also include the well-known
Casimir force \cite{Casimir1948} that has been studied, e.g., between a metallic
sphere and a nanostructured surface \cite{Mohideen1998,Intravaia2013}.
Historically, and rather confusingly, the nonretarded regime with
$U_{\text{vdw}}(z)=-C_3/z^3$ is sometimes called the van der Waals (vdW)
potential in order to distinguish it from the retarded (or Casimir-Polder) regime,
which asymptotically converges to $U_{\text{ret}}(z)=-C_4/z^4$ at a large
distance from the surface $z$. In the current usage, the Casimir-Polder interaction consistently refers to the dispersion interaction between a microscopic (atom or molecule) and a macroscopic object independent of the distance regime.

Pioneering work with Rydberg atoms \cite{Sandoghdar1992} predominantly probed
the nonretarded regime even at atom-surface distances as large as 1 $\mu$m due
to major contributions from atomic transitions in the mid-IR. On the other
hand, when the atomic transitions are in visible or near UV regions --
such as for atoms in their ground states -- the atom-surface interaction
will be in the CP regime. This scenario is relevant for ground-state atomic
beams \cite{Sukenik1993}, cold atoms near atomic mirrors \cite{Landragin1996}
and quantum reflection \cite{Shimizu2001}.
Very few experiments thus far have studied the crossover regime where neither
limit holds, typically using an adjustable repulsive dipolar force
\cite{Bender2010}. Studying atom-surface interactions with reasonable accuracy
is of major importance as these fundamental fluctuation-induced interactions
have not been yet measured with an accuracy better than 5-10\% whatever
the experimental approach.

In this Letter, we present our experimental and theoretical investigations of
matter-wave diffraction of metastable argon atoms by a transmission nanograting
at atom-surface distances below 51 nm. The geometric constraint on the
atom-surface distance provided by the two adjacent walls is a major asset that
eliminates the quasi-infinite open space over a single surface, similar to an
ultrathin vapor cell \cite{Fichet2007}. Atom-surface interactions have
previously been studied using transmission nanogratings with atoms at thermal
velocities \cite{Grisenti1999,Lonij2009}. This Letter shows that lowering the
atomic beam velocity below 26 \ms opens up new experimental opportunities due
to larger interaction times, and produces diffraction spectra dominated by the
atom-surface interaction \cite{Nimmrichter2008}. The precise control of nanograting geometry and experimental parameters related
to the atom beam leads us to observe the minute influence of retardation. This
paves the way to accurate CP potential measurements that can be compared
against detailed theoretical models.

\begin{figure}[ht]
\includegraphics[width=0.99\linewidth]{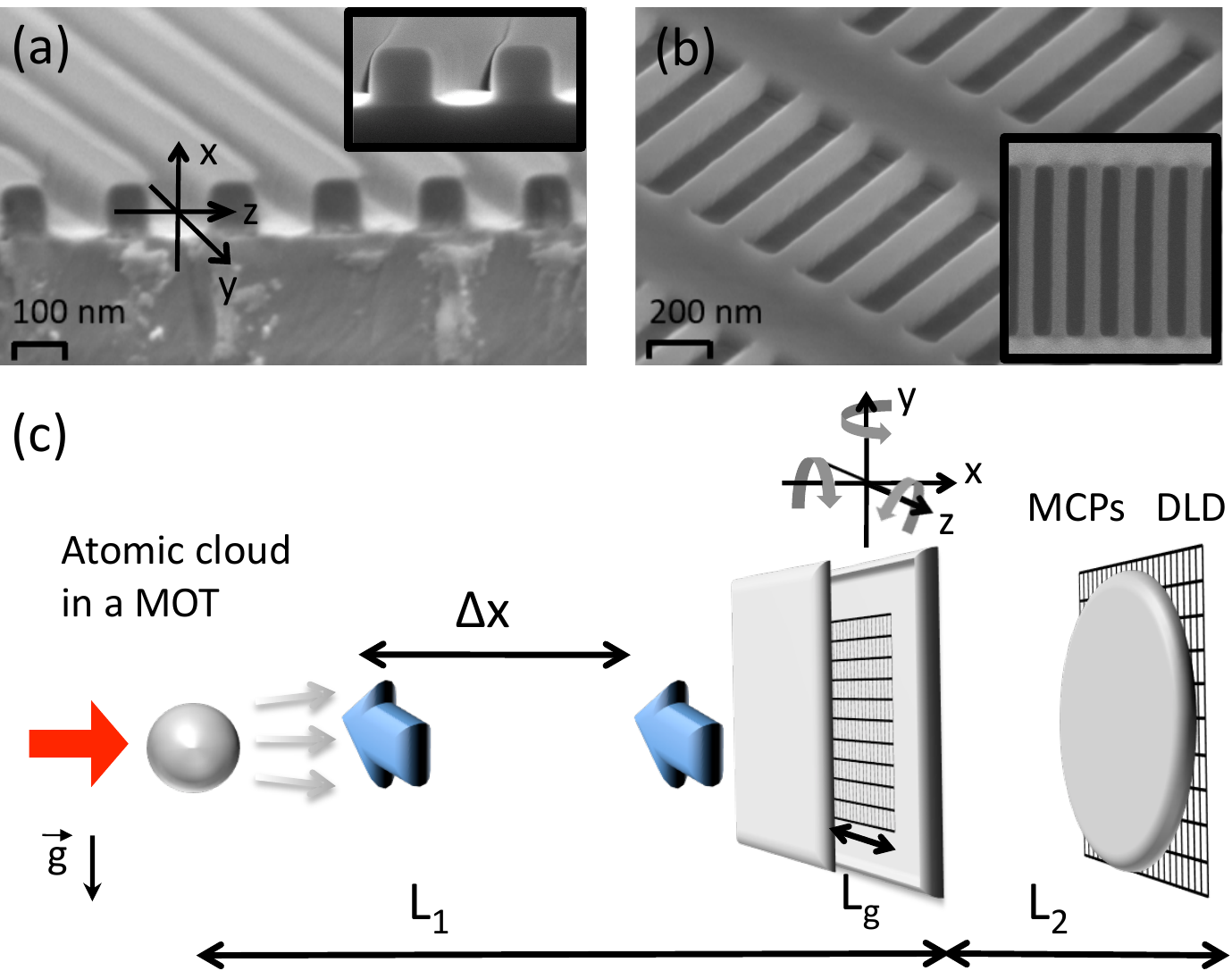}
\caption{SEM images of (a) a cleaved nanograting on a substrate and (b) a free
standing membrane. (c) Experimental setup. An atomic cloud is periodically
pushed by a laser (red arrow) from a MOT at a chosen velocity through a nanograting $L_1=56$ cm away. A time-position detector is located $L_2=308$ mm behind the nanograting. A light chopper (blue arrows) is available for accurate velocity measurement. A 6D piezo system controls the grating position.}
\label{figsetup}
\end{figure}

Transmission gratings etched into a 100 nm thick silicon nitride
(Si$_3$N$_4$) membrane are commonly made using achromatic interferometric
lithography \cite{Savas1995} using UV light, resulting in gratings with
pitch down to 100 nm covering areas of several mm${}^2$. For its versatility in
nanograting design, we chose electron beam lithography to pattern a new
generation of resists with high selectivity during etching and low line edge
roughness \cite{Thoms2014}. A 200 nm-period transmission nanograting has been
fabricated on a 100 nm thick membrane of $1\times1$ mm$^2$ in size. The
combination of 100 keV e-beam lithography and anisotropic reactive ion etching
ensures parallel walls with deviations smaller than 0.5 degrees from the
vertical \cite{Kruckel2017,Kaspar2010,SuppMat}. The SEM image of a cleaved nanograting
shows smooth and anisotropically etched walls, rounded with a 21 nm
radius of curvature along the atom propagation $x$ axis [Fig.
\ref{figsetup}(a) and inset]. Statistical analysis of SEM images reveals
a slotted hole geometry of the slits (with straight section along 90\% of the
total slit length), with a main width $W=102.7\pm0.3$ nm and a FWHM
distribution of 7 nm [Fig. \ref{figsetup}(b)].

Hybrid experiments at the interface between atomic physics and nanoscience
often use alkali atoms for convenient laser cooling and manipulation.
However, nanostructures chemically react with residual vapor
that alters the CP potential unpredictably \cite{McGuirk2004}. Using noble gas
atoms in a metastable state prevents chemical damage on nanostructures while
retaining the ability for laser manipulation. The ($3p^54s;\,^3P_2$), metastable
state of argon is used for an efficient and accurate time-position detection by
microchannel plates with 80 mm diameter in front of a delay line detector
(DLD80 from RoentDek Handels GmbH).
The experiment starts with a supersonic beam of argon followed by a
counterpropagating electron gun, which provides a flux of $10^8$
Ar$^*$ atoms per second. The cycling transition \mbox{($3p^54s; \,^3P_2$)
$\longleftrightarrow {}$($3p^5 4p;\,^3D_3$)} (with $\Gamma=2\pi\times 5.8$ MHz, $\lambda =
811.531$ nm) is used by a Zeeman decelerator to trap atoms in a magneto-optical
trap (MOT). The trap consists of anti-Helmholtz coils providing the
magnetic field and three retroreflected laser beams,  red-detuned by
\mbox{2$\Gamma$}, with \mbox{7 mW} total power and \mbox{2.54 cm} beam diameter.
Approximately $10^{7}$ atoms are trapped at a temperature of $\approx 20$ mK.

An initial pulse sequence pushes an atom cloud at a chosen velocity
orthogonally to the incoming supersonic beam at 13 Hz repetition rate
\cite{Taillandier2016}. During this time, the magnetic field remains constant
and the molasses laser beams are switched off. Simultaneously, the circularly polarized pushing laser beam (5 mW, frequency adjustable on the same cycling transition) is
turned on for 0.4 ms toward the detector and focused to 20 cm after the MOT
position. Atoms remain in the $m=+2$ state without any influence on the
diffraction process. A TOF measurement is then performed, with
the time-position detector 86.8 cm away from the MOT. With this pushing
technique, the relative spread of the atomic velocity distribution, $\Delta
v/v$, is already less than 10\%. Moreover, a time selection of 1 ms is applied
to obtain an even narrower velocity distribution. Additionally, for an absolute
velocity determination, we used a light chopper with two resonant lasers of 1
mm diameter perpendicular to the atomic beam axis, separated by $\Delta
x=266.5\pm 1.3$ mm and time triggered with a time sequence accuracy below 50
$\mu$s. We obtained mean velocities and respective uncertainties of $19.1 \pm
0.2$ \ms and $26.2 \pm 0.1$ \ms for both recorded spectra.

The vertical $y$ axis [see Fig.~\ref{figsetup}(c)] has been chosen for the
slit alignment in order to impose a diffraction expansion perpendicular to
the Earth's gravitational field. At 56 cm from the MOT, the atomic beam
diameter is much larger ($\approx5$ cm) than the entire nanograting surface and
all slits along the $y$ axis contribute equally to the signal. This
is not the case along the $z$ axis (diffraction axis) where the angular beam
distribution acts as an incoherent source and smears the signal out, in
particular interference orders that are separated by more than
2.6 (1.9) mrad for 19.1 (26.2) \ms. A compromise between atomic flux through
the nanograting and fringe visibility (smearing) is achieved with a free
opening of \mbox{$L_g=306\pm5$ $\mu$m} between the vertical
edge of the plate and the nanograting boundary. The atomic beam divergence,
$\Delta_\theta^{\text{beam}}$, through the 306 $\mu$m slit at a distance $L_1$
from the MOT fits a Gaussian profile with 1.4 mrad FWHM. As a consequence, the
beam divergence alters perceptibly the measured diffraction spectra, but in a
controlled way. The nanograting is fixed on a 6D piezo system (SmarPod 11.45
from SmarAct GmbH) to ensure a $90.0\pm 0.1$ degree angle between the atomic
beam and the $z$ axis on the nanograting. An angular deviation as
small as 0.2 degrees introduces noticeable asymmetry of the intensities of the
diffraction orders.

\begin{figure}[ht]
\includegraphics[width=0.98\linewidth]{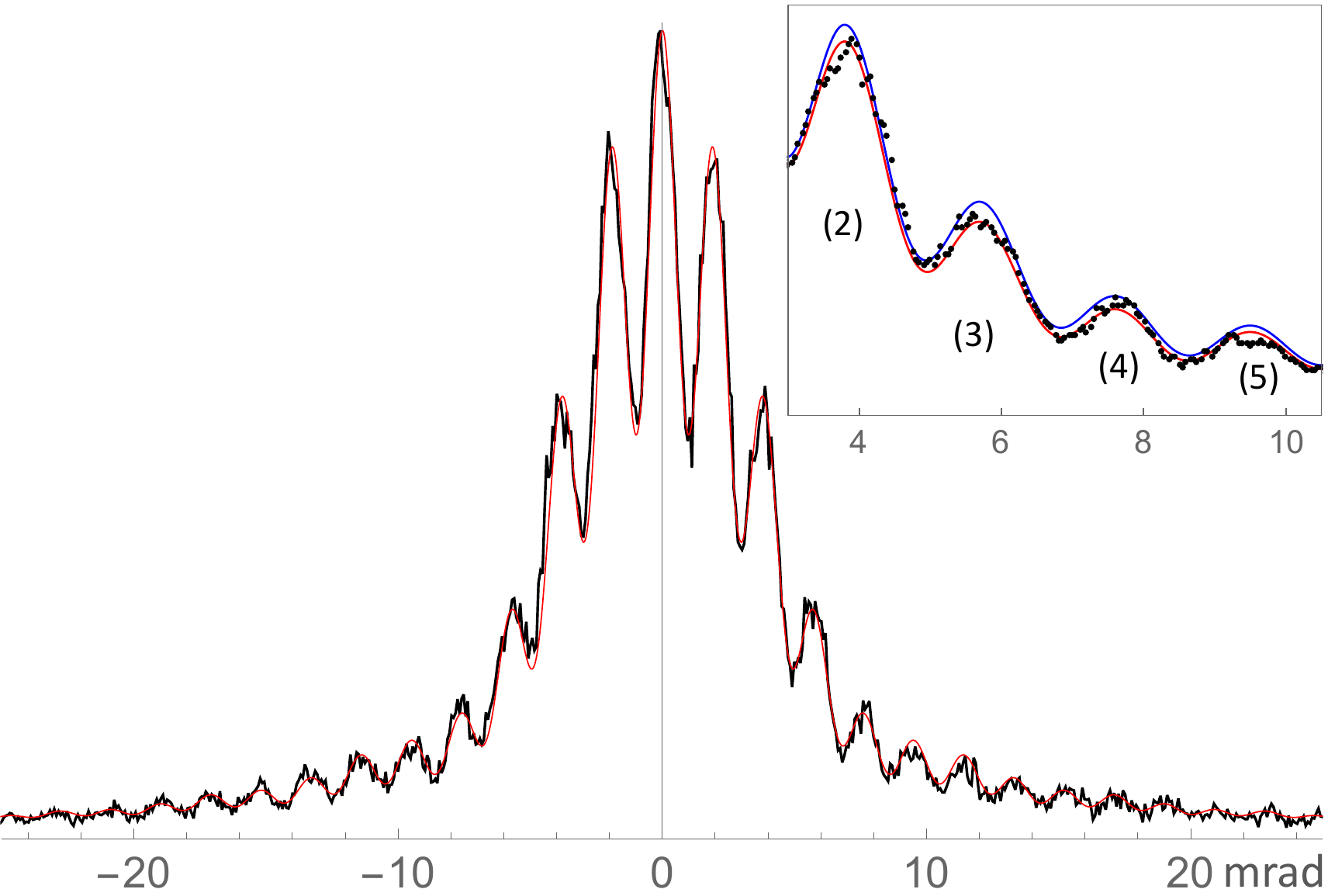}	
\includegraphics[width=0.98\linewidth]{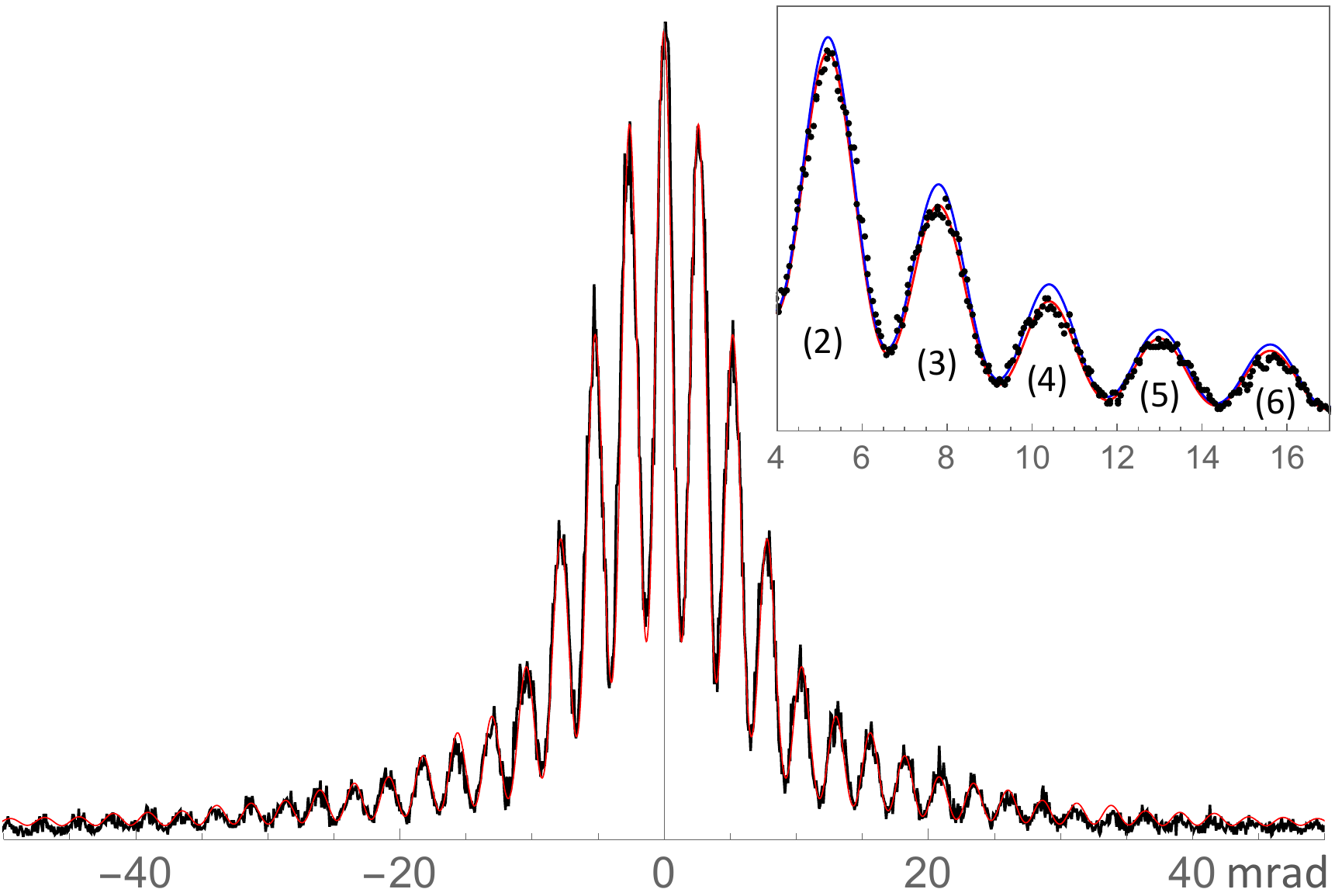}
\caption{
Experimental diffraction spectra for beam velocities of 26 \ms (top) and 19.1
\ms (bottom) in black. The red curves are theoretical spectra with a single
adjustable parameter ($d_0$). The insets show individual diffraction
orders. Black dots result from experimental spectra averaged over positive and
negative diffraction orders. Red (blue) curves are calculated with CP
(vdW) potentials.}
\label{figspectra}
\end{figure}

Two experimental diffraction spectra are shown in Fig.~\ref{figspectra} for
contributing velocities between $18.7-19.5$ \ms and $25.5-
26.9$ \ms. Small electronic aberrations in position have been corrected by the
use of a 2600-hole grid pattern, and no intensity inhomogeneity has been
noticed at the experimental accuracy level. The recording times were
40 and 13 hours for $1.006\times 10^5$ and $1.512\times 10^5$ atoms,
respectively. In general, for matter waves propagating through a transmission
grating, the diffraction spectrum envelope is determined by the wavelength and
the single slit width, while the interference peak visibility stems from the
transverse coherence length of the source. In the present situation, the
nanograting slit width effectively narrows due to atoms that are close enough
to a surface being mechanically attracted and deflected by the Casimir--Polder
potential. Metastable atoms colliding with the surface at room temperature are
scattered randomly at high velocity, and will return to their ground state
\cite{Hagstrum1954}.

For atoms at thermal velocities, the region near the surfaces the atoms needed
to avoid was assumed to be a few nanometers \cite{Grisenti1999, Lonij2009}.
Using classical trajectories, we can estimate a lower
limit for the distance at which the atoms can pass the nanobars to be
$d_0=16.2\,(14.2)$ nm at a beam velocity of $19.1\,(26.2)$ \ms or, equivalently,
an effective slit width of $W_\mathrm{eff}= 70.3\,(74.3)$ nm. Such a major
reduction cannot be neglected when explaining experimental
diffraction spectra ($2\lambda_\mathrm{dB}/W_\mathrm{eff}=14$ mrad at
19.1 \ms). The effective slit width and the exerted CP forces elsewhere in the grating explain the overall broadening of the spectra compared with an equivalent optical picture with first zeros at 5 mrad. The fringe visibility depends only on the transverse coherence
length of the atomic beam, $L_c=\lambda_\mathrm{dB}L_1/a$, with $a$ the
diameter of the incoherent source, as given by the van Cittert--Zernike theorem
\cite{Born}. However, the quadratic dispersion relation for matter waves
suppresses the dephasing compared with light \cite{Taylor1994}, and hence enlarges
$L_c$. From the cloud size in the MOT, \mbox{$a=330\pm40\ \mu$m},
followed by thermal expansion, one finds \mbox{$L_c=560\pm 45$} ($380\pm 30$) nm
at 19.1 (26.2)~\ms beam velocity.

The Huygens-Kirchhoff principle can be utilized for atoms propagating in a
potential that is small compared with their kinetic energy \cite{Feynman1948,
Kholodenko2018}. This can be justified with the help of the effective slit
approximation, which removes atoms with potential energies that are too large.
Additionally, the detection in the far field validates the
Fraunhofer approximation ($x\gg W_\mathrm{eff}^2/\lambda_\mathrm{dB}$), in which
the diffraction pattern results from the sum of wave path differences at
the nanograting output. The CP potential is included in the wave propagation as
an additional phase $\Phi_{\text{CP}}(z)$ that depends on the atom-surface
distance inside the nanograting slit $z$. In short, the total phase
can be written as $F(z,\theta)=kz \sin\theta+\Phi_{\text{CP}}(z)$ for a
detection angle $\theta$ and a wave number $k$. The incoming Gaussian wave
packets have a standard deviation $\sigma_{\text{coh}}=L_c/2$
\cite{Zecca2007, Thompson1957}. The experimental value for $L_c$ covers up to
seven slits coherently and hence, the beam cannot be considered as a plane
wave. The diffraction intensity then reads as
\begin{equation}
I(\theta)=\left| \sum_{\text{slits}} \int_{W_\mathrm{eff}} \exp\left[i
F(z',\theta)\right]  \exp\left[-\frac{z'^2}{2\sigma_{\text{coh}}^2}\right] dz'
\right|^2 .
\label{I7}
\end{equation}

In eikonal approximation, the phase shift imprinted by a potential $U_{\text{CP}}(z)$
is the integral of the potential along the particle trajectory \cite{Landau}.
Neglecting the surface potential outside the grating, one finds
\begin{equation}
\Phi_{\text{CP}}(z)=-\frac{1}{\hbar v} \int_{0}^{L} U_{\text{CP}}(x,z) dx \label{PhiCP}
\end{equation}
where $L=100$ nm is the nanograting depth and $v$ the beam velocity. However,
it is necessary to account for the exact shape of the grating along the
$x$ axis, because the slit widths increase near the output. We incorporate this
effect by using a smaller slit thickness of $L_\mathrm{eff}=95$~nm.

In pioneering experiments with nanogratings and supersonic beams
\cite{Grisenti1999, Perreault2005}, the CP potential is modeled in the
nonretarded regime as $U_{\text{vdw}}(z)=-C_3^{\text{nr}}/z^3$ for the two
adjacent surfaces. Semi-infinite surfaces are implicitly considered everywhere
inside the grating even if this is not correct near the edges.
Further, the effect of multiple reflection from the bar opposite has been
neglected \cite{Dutier2003}. These approximations are justified for fast atoms
or molecules as the overall phase shift remains small on average
\cite{Brand2015}. In the nonretarded regime, the coefficient $C_3^{\text{nr}}$
is given by the Lifshitz formula \cite{Lifshitz1958,Mavroyannis1963}
\begin{equation}
C_3^{\text{nr}}=\frac{\hbar}{16\pi^2 \epsilon_0} \int_0^\infty \alpha(i\xi)
\frac{\epsilon(i\xi)-1}{\epsilon(i\xi)+1}d\xi
\end{equation}
where $\alpha(i\xi)$ denotes the atomic dynamic polarizability,
and $\epsilon(i\xi)$ is the surface permittivity taken at imaginary
frequencies $i\xi$. For metastable argon in the $^3P_2$ state, we obtain
$C^{\text{nr}}_3=1.25$ a.u., or 5.04 meV nm$^3$ in front of a
Si$_3$N$_4$ surface with spectral responses in the UV and IR \cite{Philipp1973}.
An estimate of the uncertainty of this value is rather difficult to obtain.
First, the material may show imperfections with regard to its fabrication and
size \cite{Zollner2001} that alter the index of refraction. Second, the
electronic structure of argon in the $^3P_2$ metastable state, due to its
11.5 eV internal energy, makes the potential more sensitive to the material
optical response in the visible and near-UV regions, where accurate optical
response data are difficult to obtain. Third, the electronic core contribution
has been estimated to be in the range of 0.03 a.u. using the first ionic state.
Such a small core contribution is a clear theoretical advantage of metastable
argon compared with heavier (alkali) atoms \cite{Derevianko1999}. The temperature
dependence of $C^{\text{nr}}_3$ can be safely neglected here, in contrast to
situations in which dominant transitions appear in the mid-IR region
\cite{Laliotis2014}. Altogether, we conservatively estimate a 10\% uncertainty
that is similar to other CP calculations.

\begin{figure}[ht]
\includegraphics[width=0.99\linewidth]{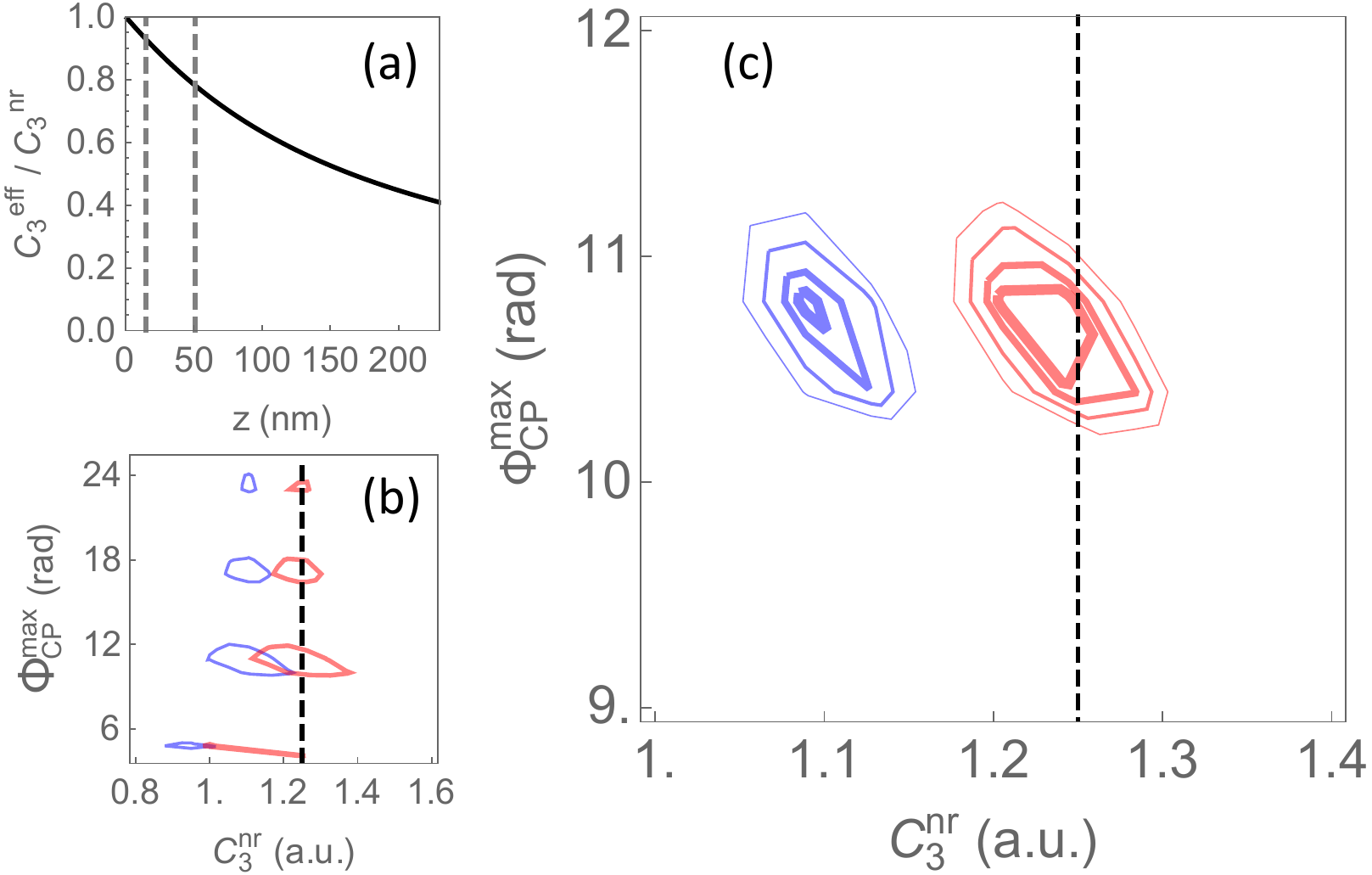}
\caption{(a): $C_3^{\text{eff}}/C_3^{\text{nr}}$ as function of atom-surface
distance. The two dashed lines represent the recorded atom-surface distance range. (b),(c): $\chi^2_{min}+n\sigma$ [$n$=40 (b), $n$ =1, 3, 6, 9 (c)] as
functions of $C_3^{\text{nr}}$ and $\Phi_{\text{CP}}^{\text{max}}$ at 26.2 \ms for
vdW (blue) and CP (red) potentials. The dashed line shows
the theoretical value for $C_3^{\text{nr}}$.}
\label{figChi2}
\end{figure}

The improvement in experimental accuracy allows us to discern the nonretarded
regime and the onset of retardation effects, that arise due to the finite field
propagation time between atom and surface. This effect becomes relevant at
distances larger than $\lambda_\mathrm{opt}/(2\pi)$.  Without resorting to a
full calculation of the exact shape of the potential, one can resort to
sophisticated interpolations. Here, we use a model derived in Ref.~\cite{Wylie}
that is built on a single atomic transition (here, $\lambda_\mathrm{opt}=811.5$
nm for Ar $^3P_2$), for which the effective coefficient
$C^{\mathrm{eff}}_3(z)$ reads as
\begin{equation}
\begin{split}
C_3^{\mathrm{eff}}(z)=C_3^{\mathrm{nr}} \left[\zeta + (2 - \zeta^2) f_1(z) + 2
\zeta f_2(z)\right] /\pi
\label{C3Wylie}
\end{split}
\end{equation}
with $\zeta(z) = 2 z (2\pi)/\lambda_\mathrm{opt}$ and $f_i(z)$ given in
\cite{f1f2}. This expression provides an interpolation between
$U_{\mathrm{vdw}}\propto z^{-3}$ at short distances and
$U_{\mathrm{ret}}\propto z^{-4}$ for $z\gg \lambda_\mathrm{opt}$. Inside the
grating, $C_3^{\text{eff}}(z)$ goes from $C_3^{\text{nr}}$ at $z=0$ to
$C^{\text{eff}}_3(51\ \text{nm})=0.78\ C_3^{\text{nr}}$ [Fig.~\ref{figChi2}(a)].
On average, the detected atoms will have experienced
$\langle C_3^{\text{eff}}\rangle=0.85\ C_3^{\text{nr}}$, which corresponds to a
15\% deviation from the nonretarded regime. This model is in very good
agreement with the complete QED calculation \cite{Scheel2008} to within a few
percent for semi-infinite surfaces, and arguably much faster to calculate.

As first shown in Ref.~\cite{Lonij2009}, the fitting procedure is extremely
sensitive to the grating geometry, and there is no unique relation between the
experimental spectrum and the set of possible theoretical parameters.
A chi-squared test is used with $\chi^2=\sum_\theta
(I^{\text{exp}}_\theta -I_\theta^{\text{theo}})^2/\sigma_{\theta,\text{exp}}^2$, where
$\sigma_{\theta,\text{exp}}$, $I^{\text{exp}}_\theta$ and
$I_\theta^{\text{theo}}$ are the experimental noise standard deviation, and the
experimental and theoretical intensities at the angle $\theta$, respectively.
Indeed, we find a linear relation between $C_3^{\text{nr}}$ and slit width as
well as nanograting thickness $L_{\text{ng}}$ to within 10\% of their nominal values:
$\Delta W=\pm$1 nm $\rightarrow \Delta C_3^{\text{nr}} =\pm$ 0.07 a.u. and
$\Delta L_{\text{ng}}=\pm10$ nm $\rightarrow \Delta C_3^{\text{nr}}= 0.16$ a.u. On the
other hand, $\sigma_{\text{coh}}$ and $\Delta_\theta^{\text{beam}}$ are
not linked to any other parameters. The remaining important parameter, $W_\mathrm{eff}=W-2d_0$ in Eq.~(\ref{I7}),
is the maximum additional CP phase shift
with $\Phi_{\text{CP}}^{\text{max}}=\Phi_{\text{CP}}(d_0)$. Note that
$\Phi_{\text{CP}}^{\text{max}}$ is different for both models as $d_0$ increases with
larger $C_3^{\text{nr}}$. Also, at larger velocities this parameter was
not considered to be critical \cite{Bender2010}.

With this, Figure~\ref{figChi2}(b) shows ($\chi^2_{\text{min}}+40\sigma$) surfaces of $C_3^{\text{nr}}$ and $\Phi_{\text{CP}}^{\text{max}}$ at a beam velocity of $26.2$ \ms and where $\sigma=6.2$ is the standard deviation for two parameters \cite{numrec}. The chosen magnification of 40$\sigma$ reveals four local minima for $\Phi_{\text{CP}}^{\text{max}}$ that have previously not been discussed, but which are necessary for a correct analysis using the Kirchhoff approximation. They correspond, respectively, to distances $z\approx 17.5, 13.1, 11.0, 10.1$ nm from the surface. We chose the global minimum for further discussion given by $\Phi_{\text{CP}}^{\text{max}}= 10.5$ rad. The dashed line is the calculated expected $C_3^{\text{nr}}$. Figure~\ref{figChi2}(c) is an enlargement of Fig.~\ref{figChi2}(b) for 1 standard deviation that rejects the van der Waals approximation (blue) by more than $30\sigma$ with regard to the calculated $C^{\text{nr}}_3$. In addition, $\chi^2$ minima for both velocities are found to be smaller with the full CP model. For further clarity, the insets in Fig.~\ref{figspectra}
emphasize the influence of both models on the spectra
calculated at $\Phi_{\text{CP}}^{\text{max}}=10.5$ rad with the theoretically expected $C_3^{\text{nr}}$. Our extracted value of $C_3^{\text{nr}}=1.24\pm 0.15$ a.u. is strongly dominated by grating geometry uncertainties ($\pm 0.13$ a.u.), where statistical error bars -- assuming a known grating -- represent only $\pm 2$\% ($\pm 0.025$ a.u.) at one $\sigma$. Such an unprecedented value is clearly connected to the ultra large diffraction spectra broadening. Note that it corrects the rather crude approximation given in Ref.~\cite{Karam2005}.

The advantage of using a slow atomic beam with a well-defined velocity rather
than a thermal beam derives from the fact that the atom-surface interaction
potential can be probed to a much higher accuracy. However, this presents
another difficulty: the Kirchhoff approximation stipulates that the (change of
the) transverse wave number has to be small compared with the longitudinal
wave number, $k_{\text{perp}}/k\ll 1$ \cite{Arsenovic2005}. The slower the atoms
become, the more (relative) transverse momentum they accumulate whilst
traversing the grating. At the cutoff that determines the effective slit
width, we can estimate the relative change in transverse wave number to be
roughly 5\% for a beam velocity of $26.2$ \ms, but already 10\% for a velocity
of $19.1$ \ms. This implies that the Kirchhoff approximation can no longer be
relied upon at slower beam velocities, at which a more detailed theoretical
description is required.

In conclusion, we have demonstrated the importance of the retarded
Casimir-Polder potential for the diffraction of metastable Ar in a range of
atom-surface distances as small as $\approx15-51$ nm with a Si$_3$N$_4$
transmission nanograting. Because of atomic velocities as slow as 19.1 \ms as well
as an accurate geometrical characterization of the nanograting, we were able to
discriminate a difference in the CP potential as small as 15\%. For lower
velocities or smaller slit widths, the semiclassical model utilized for the
simulations should be replaced by a quantum mechanical
model. Such a theoretical refinement will introduce quantum reflexion at the
slit walls and may produce, in some geometry, gravity Q-bounces as found for
neutrons \cite{Jenke2011}. This Letter opens the opportunity for unprecedented and
accurate CP potential measurements by controlling the tilt of the nanograting,
which, combined with tomography methods, would lead to a thorough understanding
of atom-surface interactions with implications for theoretical physics as well
as nanometrology. For example, the hypothetical non-Newtonian fifth force
\cite{Antoniadis2011} could be constrained by an atomic physics experiment.
Atomic quantum random walks \cite{Karski2009} based on multipath beam splitters can be simply realized
with two or more nanogratings, and closed-loop interferometers can be made
extraordinary compact.

\acknowledgments
The authors from Laboratoire de Physique des Lasers acknowledge the "Institut
Francilien de Recherche sur les Atomes Froids" (IFRAF) for supports. This work has been supported by Region Ile-de-France in the framework of DIM SIRTEQ. The
work was partly supported by the French Renatech network. The authors
acknowledge B. Darqui\'{e} for fruitful data analysis discussions.


\begin{thebibliography}{99}
\bibitem{f1f2} $f_1(z)=Ci(\zeta)\sin(\zeta)-(Si(\zeta)-\pi/2)\cos(\zeta)$ and
$f_2(z)=-Ci(\zeta)\cos(\zeta)-(Si(\zeta)-\pi/2)\sin(\zeta)$ where $Ci(z)$ and
$Si(z)$ are the integral cosine and sine functions
\end{thebibliography}
\end{document}